\documentclass{aastex63}
\usepackage{amsmath,lipsum}
\usepackage{amssymb}
\usepackage{makecell}
\usepackage{threeparttable}

\shorttitle{Rotational Velocities of Am and Non-chemical-peculiar Stars}
\shortauthors{Qin et al.}

\begin{document}
\bibliographystyle{aasjournal}

\title{Rotational Velocities of Am and Non-chemical-peculiar Stars based on Kepler and LAMOST DR5}

\correspondingauthor{A-Li Luo \& Wen Hou}
\email{* lal@nao.cas.cn ~ whou@nao.cas.cn}

\author[0000-0001-6947-8684]{Li Qin}
\affiliation{CAS Key Laboratory of Optical Astronomy, National Astronomical Observatories, Beijing 100101, China}
\affiliation{School of Information Management \& Institute for Astronomical Science, Dezhou University, Dezhou 253023, China}
\affiliation{University of Chinese Academy of Sciences, Beijing 100049, China}

\author[0000-0001-7865-2648]{A-Li Luo$^{*}$}
\affiliation{CAS Key Laboratory of Optical Astronomy, National Astronomical Observatories, Beijing 100101, China}
\affiliation{School of Information Management \& Institute for Astronomical Science, Dezhou University, Dezhou 253023, China}
\affiliation{University of Chinese Academy of Sciences, Beijing 100049, China}

\author[0000-0003-0716-1029]{Wen Hou$^{*}$}
\affiliation{CAS Key Laboratory of Optical Astronomy, National Astronomical Observatories, Beijing 100101, China}

\author[0000-0001-7607-2666]{Yin-Bi Li}
\affiliation{CAS Key Laboratory of Optical Astronomy, National Astronomical Observatories, Beijing 100101, China}

\author{Kai-Ming Cui}
\affiliation{CAS Key Laboratory of Optical Astronomy, National Astronomical Observatories, Beijing 100101, China}
\affiliation{University of Chinese Academy of Sciences, Beijing 100049, China}
\author{Fang Zuo}
\affiliation{CAS Key Laboratory of Optical Astronomy, National Astronomical Observatories, Beijing 100101, China}
\affiliation{University of Chinese Academy of Sciences, Beijing 100049, China}
\author[0000-0003-1454-1636]{Shuo Zhang}
\affiliation{Department of Astronomy, School of Physics, Peking University, Beijing 100871, P. R. China}
\affiliation{Kavli institute of Astronomy and Astrophysics, Peking University, Beijing 100871, P. R. China}

\author[0000-0001-6767-2395]{Rui Wang}
\affiliation{CAS Key Laboratory of Optical Astronomy, National Astronomical Observatories, Beijing 100101, China}

 \author{Jin-Shu Han}
\affiliation{School of Information Management \& Institute for Astronomical Science, Dezhou University, Dezhou 253023, China}

\author{ Li-Li Wang}
\affiliation{School of Information Management \& Institute for Astronomical Science, Dezhou University, Dezhou 253023, China}

\author{Yan Lu}
\affiliation{CAS Key Laboratory of Optical Astronomy, National Astronomical Observatories, Beijing 100101, China}
\affiliation{School of Information Management \& Institute for Astronomical Science, Dezhou University, Dezhou 253023, China}
\affiliation{University of Chinese Academy of Sciences, Beijing 100049, China}

\author[0000-0001-5738-9625]{Xiang-Lei Chen}
\affiliation{CAS Key Laboratory of Optical Astronomy, National Astronomical Observatories, Beijing 100101, China}
\affiliation{University of Chinese Academy of Sciences, Beijing 100049, China}

\begin{abstract}

Rotation is a critical physical process operating in the formation of Am stars. There is a strong correlation between low-velocity rotation and chemical peculiarity. However, the existence of many non-chemical-peculiar slow rotators challenges the understanding of Am stars. The purpose of our work is to search for low-velocity rotating non-chemical-peculiar A-type stars and Am stars and to make a comparative analysis. In this paper, we pick out a
sample from the LAMOST–Kepler project, including 21 Am stars, 125 non-chemical-peculiar slow rotators, and 53 non-chemical-peculiar fast rotators. We calculate the rotational frequencies through a periodic change of light curves caused by inhomogeneous stellar surfaces and then obtain the rotational velocities. For slow rotators, the age of Am stars is statistically younger than that of non-chemical-peculiar stars in the same temperature bin. In the
comparison of the period, the average amplitude, and stellar mass of Am and non-chemical-peculiar stars, we discover that there is no difference in the photometric variability between Am and non-chemical-peculiar stars, which implies similar inhomogeneities on the surfaces. The average amplitude of non-chemical-peculiar stars has a downward trend with the increase of effective temperature and stellar mass, which is consistent with the theoretical
prediction caused by weak dynamo-generated magnetic fields in A-type stars. In addition, we confirm four nonchemical-peculiar stars that have flares by checking field-of-view images, pixel images, and pixel-level light curves.

\end{abstract}


\section{Introduction}
\label{sec:1}
The Am stars are a subclass of chemical peculiar (CP) stars, whose abundance of Fe peak and heavier elements increases and the abundance of light elements such as Ca and Sc decrease in the stellar atmospheres. Most Am stars are A- and early F-type main-sequence (MS) stars and have a low rotation velocity. The abundance anomalies in Am stars are thought to be a result of the chemical separation driven by radiative levitation and gravitational settling. However, the chemical separation in normal A stars is countered by the effects of rotation. So the slow rotation is a key factor operating in the Am stars.

Previous studies observed that the rotational velocities of A stars show a bimodal distribution \citep{1965ApJ...142.1594C,1995ApJS...99..135A} and explained the slowly rotating population should be Am or Ap stars. However, some normal A-type stars have been found to have slowly rotating velocities \citep{2019MNRAS.487.4695S,2020MNRAS.498.2456S,2020MNRAS.492.3143T},  which challenges our understanding of Am stars. Therefore, it is necessary to conduct a comparative analysis of the Am stars and the low-velocity non-CP rotators. 

Since the hot A-type Vega star was discovered to have a weak magnetic field \citep{2009A&A...500L..41L,2014A&A...568C...2P,2017MNRAS.472L..30P}, some studies have also successively discovered weak magnetic fields on four Am stars: the Sirius A ($0.2\ \pm\ 0.1\ \rm{G}$) \citep{2011A&A...532L..13P}, Alhena ($\sim-5\ \pm\ 3\ \rm{G}$), $ \beta $ UMa ($-1\ \pm\ 0.8\ \rm{G}$), and $ \theta $ Leo ($-0.4\ \pm\ 0.3\ \rm{G}$) \citep{2016MNRAS.459L..81B, 2016A&A...586A..97B}. Using Zeeman-Doppler imaging technology, \citet{2015A&A...577A..64B} reconstructed the magnetic topological structure for the Vega, which contains a polar magnetic region and several low-latitude starspots on its surface, and provided strong evidence that A-type stars can show surface structures. \citet{2019ApJ...883..106C} also theoretically predicted that A-type stars have magnetic fields of $ 1-10\ \rm{G} $, which supports the existence of weak starspots on the surface. 

Using high-precision and long-term photometric data provided by the Kepler project, \citet{2013MNRAS.431.2240B, 2015MNRAS.448.1378B, 2017MNRAS.467.1830B} found that the light curves of $\sim $ 44\% A-type stars display rotational modulation features. \citet{2020MNRAS.498.2456S} presented a high-resolution spectroscopic study of MS A and revised the incidence rate to $\sim $ 30\%. Although Ap stars are known to exhibit photometric variabilities caused by chemical patches (spots) on the stellar surface, they account for only $\sim$ 10\% of A-type MS stars \citep{2019MNRAS.483.2300S}. Obviously, about 20\% of non-Ap stars should also have photometric variabilities in A-type MS stars. \citet{2020MNRAS.492.3143T} carried out further studies on A and Am stars with `hump and spike' characteristics, and proved again that there are inhomogeneous surface structures on the surface of A-type stars.

In order to better understand Am stars, the following work has two research purposes. One purpose is to search for low-velocity rotating Am stars and non-CP stars, using the rotation modulation feature caused by inhomogeneous surface structures. Another purpose is to compare and analyze the various properties of Am and non-CP stars and try to obtain valuable clues about the Am phenomenon.

The paper is organized as follows. In Section 2, we use spectra to screen out Am stars and non-CP stars. In Section 3, we calculate the rotational velocity of Am and non-CP stars. In Section 4, we analyze the rotational velocity, age, temperature, period, amplitude, and flare of the Am and non-CP stars. Finally, summarize in Section 5.

\section{Selection of Am and Non-CP Stars}\label{sec:2}
The Kepler telescope has continuously observed the Kepler field of view (FOV) for 4 yr, providing high-precision long-term time-series photometric data. The Large Sky Area MultiObject Fiber Spectroscopic Telescope is a multi-fiber large-aperture reflective Schmidt telescope that can acquire 4000 spectra in one exposure \citep{2012RAA....12.1197C,2012RAA....12..723Z}. LAMOST-Kepler project is a sub-project of the LAMOST Survey, and its primary purpose is to obtain as many spectra as possible in the Kepler FOV and to provide reliable atmospheric parameters for these stars \citep{2015ApJS..220...19D}. The data provided by the project support this work.

\subsection{Data Preselecting}\label{sec:21}
Since the Am phenomenon often occurs in A-type and early F-type MS stars and the spectral information of Am stars is mainly concentrated at the blue band of spectra, we use three criteria to screen high-quality spectra from the LAMOST DR5 \footnote{\url{http://dr5.lamost.org/v1/}}. (1) The spectral types of stars are A, F0, F1, and F2. (2) The signal-to-noise ratio of the g-band is greater than 50. (3) There is no zero flux at the blue band of spectra. After removing repeated observation spectra, there are 193,345 spectra left. We adopt the template matching method to obtain the atmospheric parameters of these stars, and use $ 6500\ \text{K}\leqslant T_{\rm eff}\leqslant11000\ \text{K}$ and log $ g\geqslant4.0\ \rm dex$ to constrain the sample further, which makes 98,202 spectra left. Next, we cross match the sample with Kepler data using a radius of three arcsec and obtain 1,954 common sources. Since the resolution of Kepler is four arcsec, we finally obtain a sample of 1922 stars in the Kepler field after removing the objects with multiple sources within four arcsec.

\subsection{Selecting Am Stars}\label{sec:22}
\citet{2019ApJS..242...13Q} investigated the empirical separation curve, the MKCLASS program, and six machine-learning algorithms in order to search for Am stars. According to their performance, the random forest algorithm was adopted and  15,269 Am candidates were identified in LAMOST DR5 and 9,372 Am stars were released through manual inspection. By cross matching the preselected sample with the Am catalog of \citet{2019ApJS..242...13Q}, 111 Am stars are obtained as members of the Am group.

\subsection{Selecting Non-CP Stars}\label{sec:23}
Two groups of CP stars with spectral types of A and early F-type are Am and Ap stars. Except for Am and Ap, other A and early F stars are collectively referred to as non-CP stars. In order to avoid the contamination of CP stars to the non-CP group, we need to remove Am and Ap candidates from the sample. For Am stars, we utilize the random forest algorithm and the MKCLASS software \citep{2016AJ....151...13G} to obtain 344 Am candidates and remove them from our sample.

For Ap stars, we make use of the method of \citet{2019ApJS..242...13Q} to remove Ap stars. Generally, Ap stars are characterized by the extreme enrichment Sr, Cr, Eu, or Si elements in their stellar atmosphere. Although some Am stars also exhibit the overabundance of these elements, their abundance values rarely exceed 2.0 dex relative to the solar abundance \citep{1987ApJS...65..137L,1996Ap&SS.237...77S,2014MNRAS.441.1669C,2015MNRAS.451..184C}. \citet{2019ApJS..242...13Q} used a set of theoretical templates with [Sr/H], [Cr/H], [Eu/H], and [Si/H] = 2.0 dex as the reference templates, which are generated by the SPECTRUM program \citep{1999ascl.soft10002G}. Since $\lambda $ 4077 \AA\ is the strongest spectral line features of Ap stars \citep{2009ssc..book.....G}, \citet{2019ApJS..242...13Q} calculated the equivalent width (EW) of the observed spectra and the reference templates around $ \lambda $ 4077 \AA\ and considered the stars of which EW is larger than that of template spectrum as Ap candidates. According to this criterion, we remove 42 Ap candidates from the remaining sample.

In addition, \citet{2020A&A...640A..40H} found 1002 Ap stars in LAMOST DR4 using the characteristic of $ \lambda $ 5200 \AA\ flux depression and \citet{2016AJ....151...13G} provided an MK spectral classification catalog using the MKCLASS program for the LAMOST-Kepler subproject. By cross matching the two results with the remaining sample, we have found another 6 and 265 Ap candidates respectively, and removed them from the sample. Finally, we put the remaining stars in the sample into the non-CP group, a total of 1,154.

 \begin{figure}
 	\centering
 	\includegraphics[width=14cm]{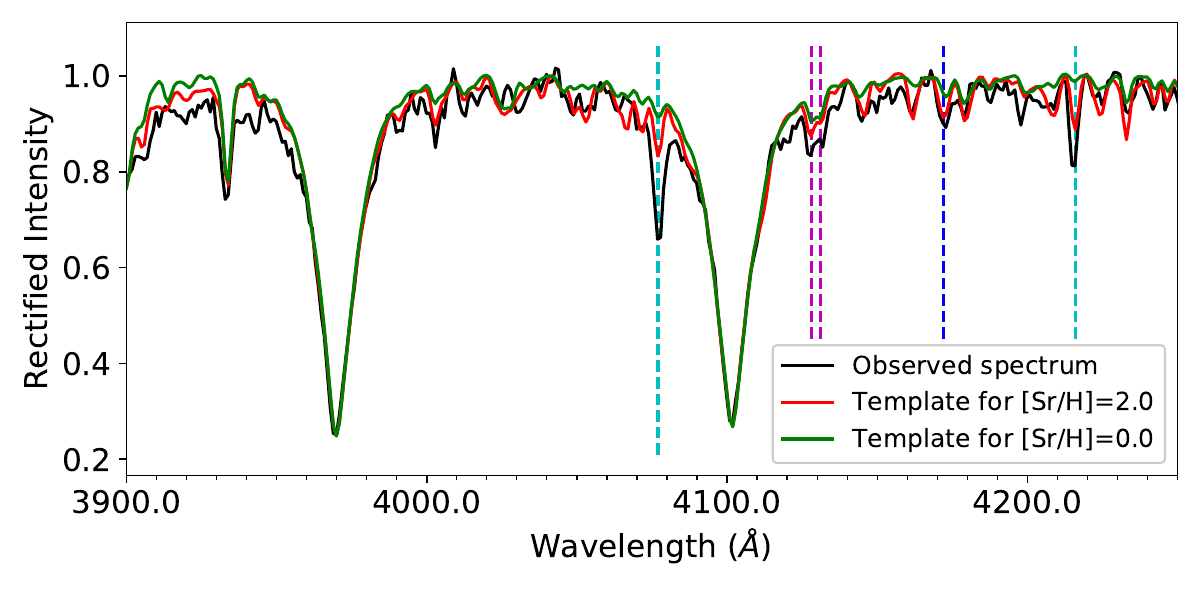}
 	\caption{The observed spectrum of an Ap star and its best matched theoretical templates with different abundances. The black, green, and red spectral lines respectively represent the observed spectrum, the matched standard template with [Sr/H], [Cr/H], [Eu/H], and [Si/H] = 0.0 dex, and the corresponding theoretical template with [Sr/H], [Cr/H], [Eu/H], and [Si/H] = 2.0 dex. The absorption lines at SrII $ \lambda $ 4077 and 4216 \AA\ are marked with cyan dashed lines. The absorption lines of SiII $ \lambda $ 4128, 4131 \AA, and CrII $ \lambda $ 4172 \AA\ are marked with magenta and blue dashed lines.}
 	\label{fig:1}
 \end{figure}

Figure \ref{fig:1} shows the normalized LAMOST spectrum of an Ap star and its normalized best matched theoretical templates with different element abundances. The black curve is a low-resolution LAMOST spectrum, and the green and red curves represent the matched template spectra with solar element abundance and enriched abundances of Sr, Cr, Eu, and Si, respectively. The cyan dashed line labels the position of the absorption lines at SrII $ \lambda $ 4077 and 4216 \AA. The absorption lines of SiII $ \lambda $ 4128, 4131 \AA, and CrII $ \lambda $ 4172 \AA\ are marked with magenta and blue dashed lines. From the spectral line of $ \lambda $ 4077 \AA, we can easily recognize that this is an Ap star.

\section{Rotation Velocity }
\subsection{Selection Criteria}\label{subsec:sele_crite}
When a star rotates, the uneven brightness areas (spots) on its surface may appear and disappear in the field of view, resulting in periodic changes in the brightness of the star.  There are two characteristics about the rotation modulation of A-type stars \citep{2011A&A...532L..13P,2011MNRAS.415.1691B}: $ f_1 < 3 \text{ day}^{-1} $ and $ f_2 = 2f_1\pm\Delta f $, where $ f_1 $ is the rotating frequency, $ f_2 $ is the first harmonic of $ f_1 $, $ \Delta f$ is the frequency uncertainty. The first criterion is based on the fact that the frequency  cannot be greater than the critical rotation frequency of zero-age MS A-type stars ($f_{crit} \approx 3 - 3.7 \text{ day}^{-1} $). The second criterion is to distinguish the frequency caused by stellar pulsation. Our research objects are located in the area where the MS zone and the classical instability zone are overlapped on the Hertzsprung-Russell diagram. For the MS stars outside the unstable pulsation zone, the fundamental frequency should be equal to the rotation frequency except for binary systems. However, the fundamental frequency may also be the pulsating frequencies for some objects in the instability zone. Therefore, we need the second condition to constrain rotational frequency. Here, we only consider the case where the fundamental frequency is the rotation frequency.
\subsection{Method}\label{subsec:data_proc}
The Kepler telescope has two observation modes: short cadence with an observation every 58.85 s, and long cadence (LC) with an observation every 29.45 minutes. The 4 yr LC data are suitable for identifying low-frequency signals. The light curve of Kepler contains two kinds of flux data: uncorrected simple aperture photometry (SAP) flux and presearch data conditioning (PDC) flux. The latter removes the instrumental effects of SAP flux and improves the SNR of the low-frequency parts. We utilize Kepler LC PDC data to search for target stars.

Most Kepler targets have 17 quarters of data ($ \text{Q}_1-\text{Q}_{17} $), and some targets also have pilot observation data ($ \text{Q}_0 $).  For each target, there is a flux jump between adjacent quarters. Therefore, before splicing the data of all the quarters of the target into a complete light curve, we divide the flux by the median flux to obtain the normalized light curve for each quarter data.

We adopt the prewhitening method to iteratively extract the significant periodic signals and corresponding frequencies in the light curve of a target star. The specific steps are as follows: (1) adopt the Lomb-Scargle (LS) algorithm to obtain the periodogram of the light curve. (2) In the periodogram, get the frequency corresponding to the maximum amplitude. (3) Make use of the $ 5\sigma $ of all amplitudes as a threshold to distinguish signal from noise. (4) If the maximum amplitude is greater than $ 5\sigma $, the corresponding period signal is subtracted from the light curve. The above four steps are repeated until there is no significant frequency or the iteration exceeds 200 times.

We apply the following Equation \ref{eq:f_cor} and Equation \ref{eq:f_chi} to fit the corresponding period signal in the fourth step \citep{2018ApJS..236...16V}.
\begin{equation}
y_{model}(t,f) = A_f^0+\sum_{k=1}^{K}A_f^{k}\textmd{sin}(2\pi kf(t-\phi_f^{k}))
\label{eq:f_cor}\\
\end{equation}

\begin{equation}
\chi^2(f)\equiv\sum_{n}(y_n-y_{model}(t_n,f))^2
\label{eq:f_chi}\\
\end{equation}

where $A_f^0$, $A_f^k$, and $\phi_f^k$ are free parameters. $t$ is the sampling time variable. $f$ is the frequency variable, designated here as the rotation frequency. $y_n$ is the photometry data at the $n$th moment. We regard the standard deviation of related parameters as their errors.

For the first extracted frequency $ f $, if $  f<3 \text{ day}^{-1} $ and there is a significant extracted frequency in $  [2f-\Delta f, 2f+\Delta f] $, the frequency $ f $ is marked as the rotation frequency. $ \Delta f $ is the half width at half maximum of the main peak in the sampling window. If there is an extracted frequency $ f_1 $ in $ [f/2-\Delta f, f/2+\Delta f] $, use $ f_1 $ as the rotation frequency and $ f$ as its first harmonic frequency, because 10 \% of the rotation frequency has an amplitude smaller than that of the first harmonic \citep{2019MNRAS.487.4695S}.

Then, we calculate the rotation velocities of the sample with the formula as follows:
\begin{equation}
{V}_{\text{rot}}=2\pi R f_{\text{rot}},
\label{eq:1}
\end{equation}
where $ V_{\rm rot} $ is the rotational velocity ($ \text{km s}^{-1} $), R is the stellar radius (km) provided by \citet{2020AJ....159..280B}, $ f_{\rm rot} $ is the stellar rotation frequency ($ \text{Hz s}^{-1} $) we calculated.

In our two data sets, the rotation velocities of a total of 590 stars can be calculated, including 69 Am stars and 521 non-CP stars.

\subsection{Decontamination}\label{sec:bin}
Previous studies have shown that about 60-70 \% \citep{1985ApJS...59..229A, 2007MNRAS.380.1064C} of Am stars are binary systems, and this proportion for A-type stars is about 35 \% \citep{2009AJ....138...28A}. Because the orbital frequencies of some binaries may also satisfy the criteria in Section \ref{subsec:sele_crite}, it is necessary to eliminate the pollution of the binaries. We cross match the sample with the International Variable Star Index\footnote{\url{https://www.aavso.org/vsx/}}, Kepler eclipsing binary catalog third revision \citep{2016AJ....151...68K}, LAMOST RV variable catalog \citep{2020ApJS..249...22T}, and the Simbad astronomical database\footnote{\url{http://simbad.u-strasbg.fr/simbad/}}, and then exclude 128 binary systems. In addition, because the pixel resolution of Kepler is 4 arcseconds, the photometric data of a star is easily contaminated by nearby stars. We utilize contamination information provided by \citet{2020AJ....159..280B} to remove 66 polluted sources in our sample. 

We use a sliding window with a window width of 200 days and a step length of 20 days and LS algorithm to obtain a series of maximum amplitudes for each light curve in our sample and utilize the coefficient of variation(VC=$ \sigma/E $, $ \sigma$: standard deviation, $ E$: mean) to measure the degree of maximum amplitude change over time. We also calculate the VC of 86 Eclipsing binaries that are the sources in the Kepler eclipsing binary catalog and removed from our sample.

\begin{figure}
	\centering
	\includegraphics[width=14cm]{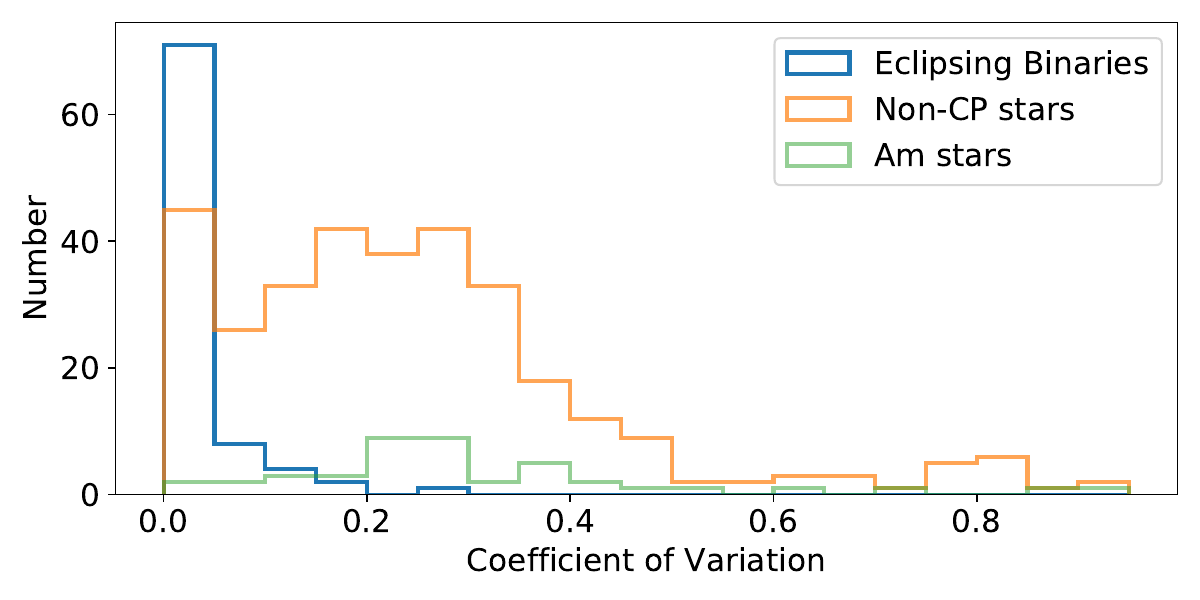}
	\caption{Distribution of the VC of eclipsing binaries, non-CP stars and Am stars with bin = 0.05. The VC of 83\% of EBs is less than 0.05. The VC values of non-CP stars show a bimodal distribution.   }
	\label{fig:CV_Amplitude}
\end{figure}

Figure \ref{fig:CV_Amplitude} is the distribution of the VC of eclipsing binaries, non-CP stars, and Am stars. The smaller the VC, the more stable the stellar light curve. EBs are mostly concentrated at the left end of the distribution, and approximately 83 \% of EBs have a variation coefficient less than 0.05. Because some EBs are affected by flares and pulsations, the shapes of their light curves are distorted. As a result, the distribution of their VC has a long tail on the right. In Figure \ref{fig:CV_Amplitude}, most of the Am stars have a VC greater than 0.05 and most of the eclipsing binary stars have a VC less than 0.05. However, the distribution of the VC of Non-CP stars presents a bimodal pattern, and the peak with the VC less than 0.05 overlaps with the peak of the eclipsing binary group. When it is impossible to determine whether a star is a binary system, removing stars with the VC less than 0.05 has two advantages:
first, they greatly reduce binary pollution;
second, we only focus on both non-CP and Am stars that show light curves with similar stability, because we hope to obtain the general difference between slowly rotating non-CP stars and Am stars. So, we removed 47 objects with a VC of less than 0.05.

\subsection{Manual Inspection}\label{sec:check}
\citet{2013MNRAS.432.1203M, 2014ApJS..211...24M} successfully adopt the Autocorrelation Function (ACF) method to detect stellar rotation periods. After calculating the ACF of the light curve of each star, we use the following formula to fit the ACF curve and obtain the optimal period value for each star. We employ the period to assist in checking the correctness of the rotation periods obtained in Section \ref{subsec:data_proc}. 

\begin{equation}
y(t)=e^{-t/\tau_{DT}}\left(Acos\left(\frac{2\pi t}{\textmd{P}_{\textmd{ACF}}}\right)+Bcos\left(\frac{4\pi t}{\textmd{P}_{\textmd{ACF}}}\right)+y_0\right)
\label{eq:DT}\\
\end{equation}

where, $\tau_{DT} $, $ \P_{\textmd{ACF}} $, $ A $, $ B $, $ y_0 $ are free parameters. $\tau_{\text{DT}} $ measures the decay timescale of periodic signals, $ P_{\textmd{ACF}} $ represents the rotation period, and $ t $ is time lags.  For details, please refer to \citet{2020MNRAS.492.3143T}.

\begin{figure*}[ht]
	\centering
	\includegraphics[scale=0.65]{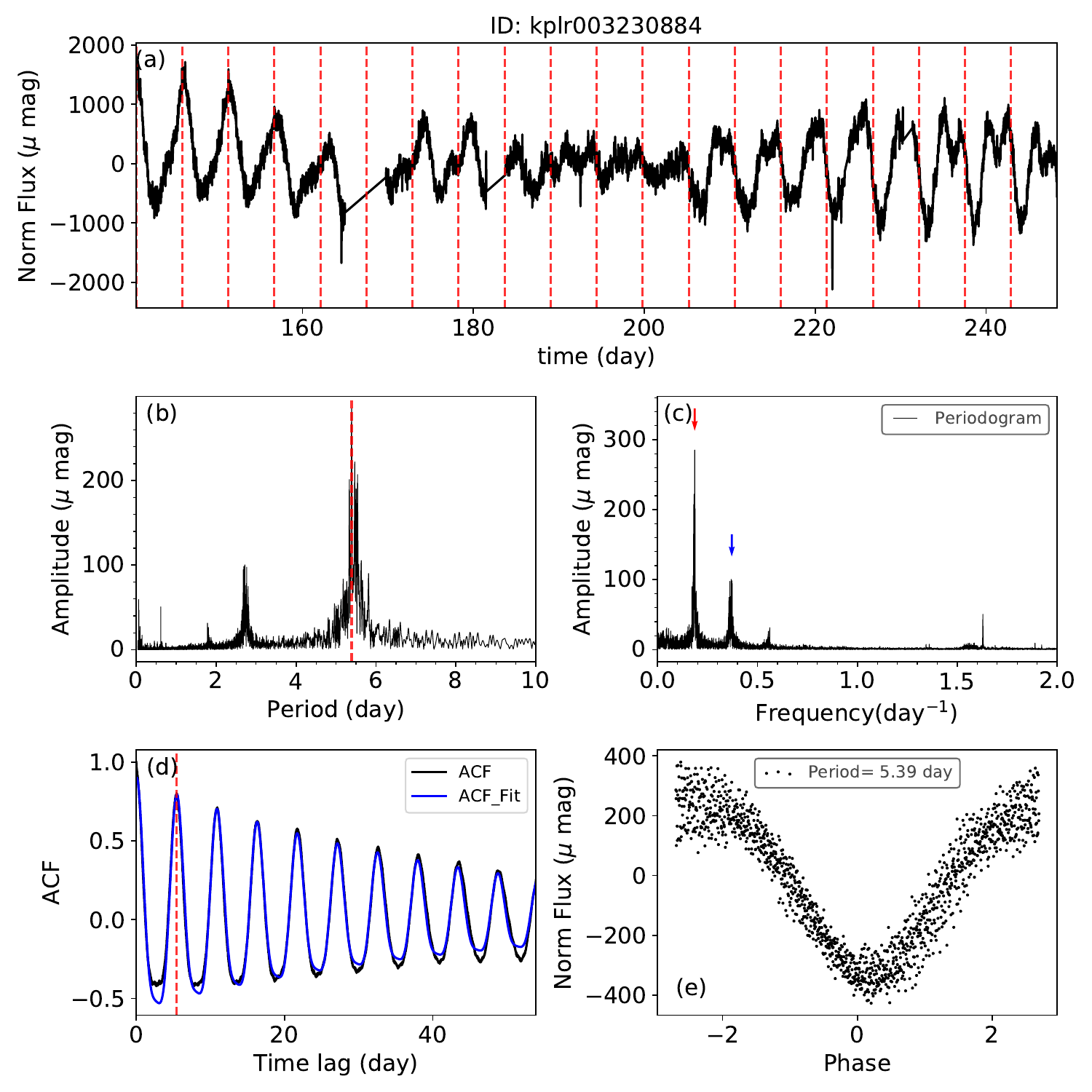}
	\caption{Legend for a manual check of rotation frequency. (a) is a segment of the light curve and the intervals of the red dashed lines indicate the rotation period we detect. (b) is the periodogram and the red dotted line is the rotation period we detected. (c) is the periodogram in frequency view. The red arrow indicates the rotation frequency, and the blue arrow indicates the first harmonic. (d) is the ACF. The blue curve is the fitted ACF. The red dashed line marks the rotation period value. (e) is folded phase. This is the case where the rotation periods provided by the two methods are consistent. }
	\label{fig:manual_chech}
\end{figure*}

Figure \ref{fig:manual_chech} is an example diagram employed in our manual inspection. Subgraph (a) is a segment of the light curve of KIC 3230884, we mark out 20 rotation periods with equally spaced red dashed lines. Subgraph (b) and (c) are periodograms provided by LS algorithm, and the $x$-axes are period and frequency respectively. The red dashed lines in (b) indicate the rotation frequency obtained by the LS algorithm. The red and blue arrows in (c) represent the rotation frequency and its first harmonic. Subgraph (d) is the ACF. The black and blue curves represent the ACF curve and the curve fitted by Equation \ref{eq:DT}. The red dashed line has the same meaning as (b). Subgraph (e) is the light curve folded on the rotation period.

In the manual inspection step, 32 objects have poor light curves or no obvious periodic changes that are removed from the sample.  For PDCSAP\_FLUX type data, to effectively remove instrumental artifacts, stellar signals with periods significantly longer than 20 days are severely suppressed \citep{2015AJ....150..133G}. Moreover, in the low-frequency band of the periodogram of many stars, the noise is severe, and some intrinsic frequencies are submerged in the noise. So as to obtain reliable rotation frequencies $ f $ to ensure sufficient confidence in the subsequent analysis, we removed 107 stars with $ f<0.05\ \text{day}^{-1} $ from the sample although it will make some slow rotators lost.

About one third of rotation frequencies are consistent with the frequencies provided by the ACF method. About a quarter of rotation frequencies is half of the frequency value provided by ACF. In the case that there is a frequency $ f_1 $ in $ [f/2-\Delta f, f/2+\Delta f] $, where $ f $ is the fundamental frequency, we adopt $ f_1 $ as the rotation frequency. However, if $ f_1 $ is not significant in the local frequency band, and the noise level of this band is far higher than the $ 5\sigma $ threshold of global noise, $ f_1 $ can not be used as the rotation frequency. Here we manually corrected these wrong judgments. For the remaining more than 100 stars, the ACF provides wrong rotation frequencies because Equation \ref{eq:DT} does not fit ACF well or fails to fit. In general, the highest peak location corresponds to the rotation period in the stellar ACF graph. In this case, we manually identify the rotation period of the ACF and then determine the final rotation periods for stars from the result of two methods.

In addition, we use the projected rotational velocity ($ v\,\text{sin}\,i$) to further check the reliability of the rotational velocity for our sample. The $ v\,\text{sin}\,i$ is measured by means of template matching. Since the resolution of the LAMOST spectrum is $R \approx 1800 $, we can only recognize  the $ v\,\text{sin}\,i$ larger than 167 km s$^{-1}$. There are six outliers in our sample, with both Vrot $<\, v\,\text{sin}\,i$ and $ v\,\text{sin}\,i$ $>=$ 167 km s$^{-1}$, and five of them are with $ T_{\rm eff} $ $>$ 11,000K according to the temperature provided by \citet{2020AJ....159..280B}. These objects are not within the scope of our research, thus we removed them from our sample.
\section{Result and Analysis}\label{sec:res_ana}
After eliminating contamination and manual inspection, we finally obtain a sample of 199 stars with reliable rotational velocity, including 21 Am stars and 178 non-CP stars. Table \ref{chartable} is the sample list. 

\begin{longrotatetable}
\begin{deluxetable*}{lllllllllll}
\tablecaption{ The Sample of non-CP stars and Am stars included in this study. \label{chartable}}
\tablewidth{800pt}
\tabletypesize{\scriptsize}
\tablehead{
\colhead{obsid\_LAMOST} &\colhead{KIC} & \colhead{R.A.(J2000)} & \colhead{Decl.(J2000)} & \colhead{$f$ (1/d)} & \colhead{err\_$f$ (1/d)} & \colhead{VC} & \colhead{Rper ($\mu$mag)} & \colhead{$err$\_Rper ($\mu$mag)} & \colhead{$V$rot (km s$^{-1}$)}  &\colhead{$E\_V$rot (km s$^{-1}$)}\\ 
\colhead{-} &\colhead{$e\_V$rot (km s$^{-1}$)} &\colhead{$T$eff (K)} & \colhead{$E\_T$eff (K)} & \colhead{$e\_T$eff (K)} & \colhead{log$g$} & \colhead{$E$\_log$g$} & \colhead{$e$\_log$g$} & \colhead{[FeH]} & \colhead{$E$\_[FeH]} & \colhead{$e$\_[FeH]}\\ 
\colhead{-} &\colhead{$M$ ($M_{\odot} $)} & \colhead{$E\_M$ ($M_{\odot} $)}& \colhead{$e\_M$ ($M_{\odot} $)}& \colhead{$R$ ($R_{\odot} $)}& \colhead{$E\_R$ ($R_{\odot} $)}& \colhead{$e\_R$ ($R_{\odot} $)}& \colhead{Age (Gyr)}& \colhead{$E$\_Age (Gyr)}& \colhead{$e$\_Age (Gyr)}& \colhead{Note}
} 
\startdata
362706149 & 10751494	&298.51474	&48.081051	&0.137767  	&1.33E-06 	&0.0900 	&1241.6396 	&311.7873 	&18.7633 &0.6412\\
 - &-0.6203 &7832.5	&190.2	&-178.9	&3.874	&0.034	&-0.036	&0.012	&0.162	&-0.18\\
 - &1.998 &0.095 &-0.097 &2.692	&0.092	&-0.089	&0.76	&0.11	&-0.1	&1\\
359915148	&9224836	&294.36871	&45.618618	&0.969524	&7.13E-06	&0.0934 	&258.1477 	&40.9170 	&97.3169 &2.9431 \\
- &-2.1582 &7399	&156.7	&-190.2	&4.058	&0.031	&-0.032	&0.008	&0.165	&-0.15\\
- &1.663	&0.091 &-0.093	&1.984	&0.06	&-0.044	&0.97	&0.27	&-0.24	&1\\
364315154	&10229037	&298.85431	&47.214458	&0.895966  	&1.03E-05 	&0.1283 	&522.2908 	&75.6655 	&102.0364 &3.3090\\
- &-2.8104 &7570.6	&178.2	&-165.8	&4.014	&0.031	&-0.031	&0.284	&0.13	&-0.157	\\
- &1.935	&0.071 &-0.081	&2.251	&0.073	&-0.062	&0.64	&0.15	&-0.13	&1\\
159605045	&6116612	&289.623292	&41.493333	&0.051161 	&1.64E-06 	&0.1296 	&974.4061 	&290.9984 	&5.7204 &0.1812\\
-  &-0.1682 &7455	&170.7	&-169.7	&4.025	&0.031	&-0.03	&0.332	&0.147	&-0.171\\
- &1.91	&0.064 &-0.073	&2.21	&0.07	&-0.065	&0.64	&0.15	&-0.14	&1\\
360011245	&9655461	&294.36267	&46.381577	&1.023298 	&1.21E-05	&0.1540 	&313.2878 	&58.5277 	&89.4610 &3.3134\\
- 	&-3.1063 &7930.9	&204.5	&-210.2	&4.19	&0.034	&-0.037	&-0.005	&0.157	&-0.174\\
- &1.706	&0.088 &-0.097	&1.728	&0.064	&-0.06	&0.55	&0.29	&-0.24	&1\\
347102079	&8866325	&283.2500	&45.17713	&0.594958 	&4.67E-06	&0.3419	&740.0420	&148.4454	&50.5690 &1.2943\\
- 	&-1.3545 &6909.8	&153.2	&-157.8	&4.127	&0.032	&-0.031	&-0.15	&0.157	&-0.133\\
- &-0.054	&1.385	&0.065 &1.68	&0.043	&-0.045	&1.65	&0.34	&-0.39	&0\\
572003004	&7499369	&280.347	&43.19934	&1.327190 	&2.55E-06	&0.0847 	&1265.4558 	&382.8620 	&102.1293 &3.0216\\
- 	&-2.8873 &7615.5	&177.8	&-173.9	&4.245	&0.032	&-0.035	&-0.144	&0.112	&-0.156\\
- &1.493	&0.074 &-0.071	&1.521	&0.045	&-0.043	&0.75	&0.36	&-0.37	&0\\
52415199	&4831769	&290.590583	&39.903583	&0.756612 	&9.13E-06	&0.4453 	&72.9952 	&38.7441 	&90.6066 &2.2967\\ 
- 	&-2.2202 &8607.8	&174.8	&-165	&3.975	&0.029	&-0.028	&-0.284	&0.148	&-0.122\\
- &1.945	&0.086 &-0.076	&2.367	&0.06	&-0.058	&0.7	&0.08	&-0.09	&0\\
52413111	&5104731	&293.362208	&40.203583	&1.149118  	&3.64E-06	&0.1710 	&85.2614 	&30.8682 	&143.4822 &3.1975 \\ 
- 	&-3.0231 &8153.8	&185.5	&-169.4	&3.937	&0.027	&-0.028	&-0.158	&0.147	&-0.143\\
- &1.935	&0.082 &-0.086	&2.468	&0.055	&-0.052	&0.75	&0.1	&-0.1	&0\\
\enddata
\tablecomments{The complete table can be downloaded in the online journal and from \url{http://paperdata.china-vo.org/Qinli/2021/Am2\_Vrot.csv}. A portion is shown here for guidance regarding its form and content. The first four columns show the ID of LAMOST spectrum, Kepler ID, right ascension, and declination of J2000 in degrees. The next three columns are rotation frequency, error of rotation frequency, coefficient of variation. Rper is the average amplitudes in light curves segmented by period. err\_Rper is the standard deviation of Rper. The next four colums are rotation velocity, upper error, lower error and projected rotation velocity. The last column is note column: the symbols 1 and 0 correspond to Am stars and non-CP stars. The rest of the parameters are from the catalog in \citet{2020AJ....159..280B}.}
\end{deluxetable*}
\end{longrotatetable}

\subsection{Rotation Velocity, Age, and Effective Temperature}\label{sec:V_M_A}

\begin{figure}
	\centering
	\includegraphics[width=12cm]{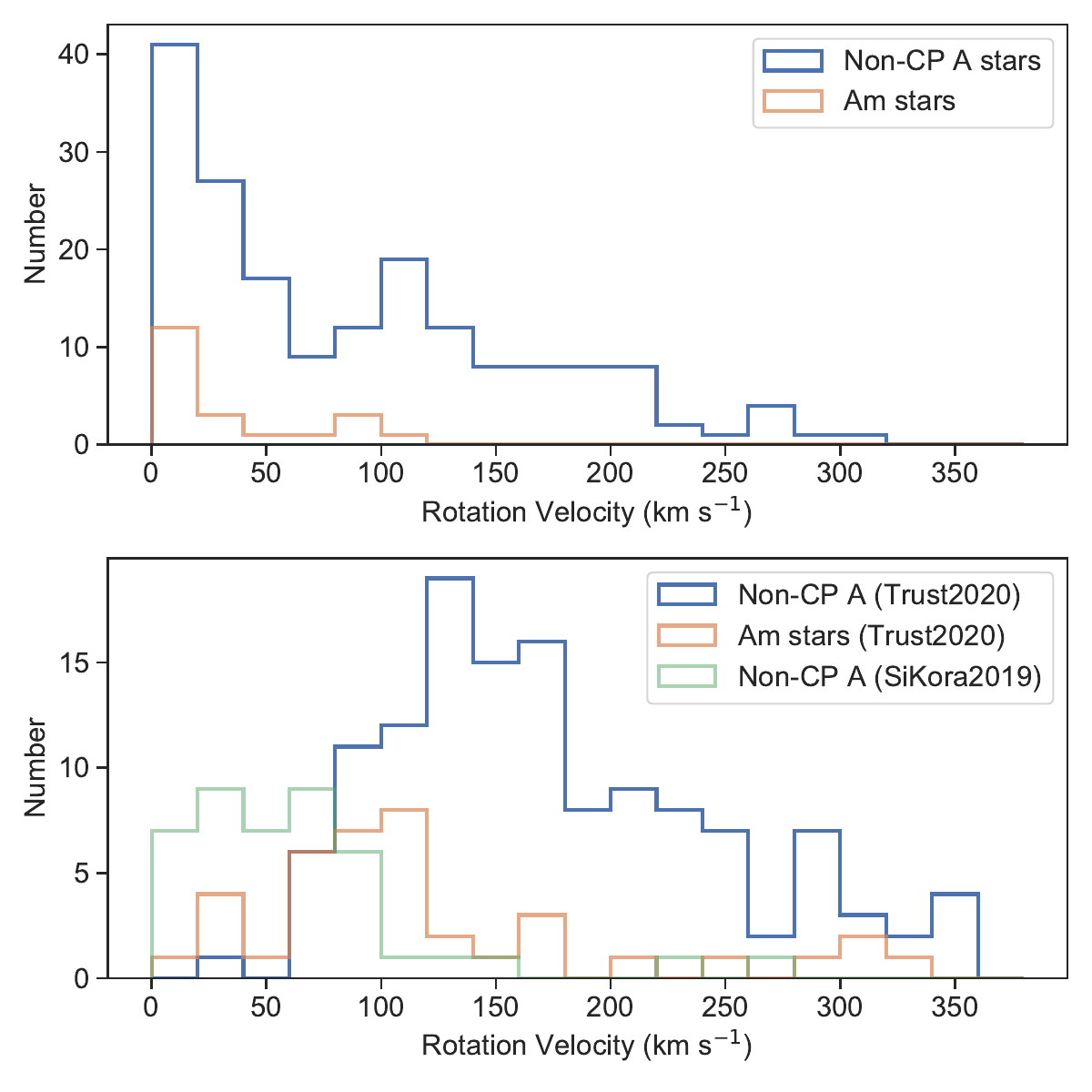}
	\caption{The distribution of rotation velocity. The top panel is the distribution of the rotation velocity of our sample. The bottom panel is the distribution of the rotation velocity from \citet{2020MNRAS.492.3143T} and \citet{2019MNRAS.487.4695S}. }
	\label{fig:vrot_M}
\end{figure}

The top panel of Figure \ref{fig:vrot_M} shows the distribution of the rotation velocity of non-CP stars and Am stars in a bin of 20 $\text{km s}^{-1}$. The rotation velocity of all Am stars is lower than 120 $\text{km s}^{-1}$, which is consistent with the results of previous studies about Am stars. However, in the rotation velocity distribution of non-CP stars, there are many low-velocity rotating non-CP stars.  The existence of those normal slow rotators will profoundly affect our understanding of the CP stars, especially for Am stars.

Recently, \citet{2020MNRAS.492.3143T} studied the rotating variable in normal A and Am stars with Kepler data. The velocity distribution of their work is shown in the bottom panel of Figure \ref{fig:vrot_M} as comparison. They search for the rotational variable with `hump and spike' features named by \citet{2018MNRAS.474.2774S}. They consider that a hump is a group of frequencies of r modes, and the spike represents the rotation frequency. The average rotation velocity of normal A stars and Am stars found by Trust are 161 and 105 $\text{km s}^{-1} $, which are both larger than our results (86.8 and 33.4 $\text{km s}^{-1} $). Because the r-mode is common in stars with medium and high-velocity rotation, it is understandable that samples with this feature have higher rotation velocity. \citet{2019MNRAS.487.4695S} utilized TESS data to identify 134 high-probability rotating variables in A-type stars, including 48 non-CP stars. Based on the radius and rotation frequency they provided, we calculated the rotational velocity of these stars. The velocity distribution is also shown in the bottom panel of Figure \ref{fig:vrot_M}. Most non-CP stars are also concentrated in the low-velocity area, similar to the distribution of our non-CP stars.

\citet{2012A&A...537A.120Z} utilized the rotation-dependent evolutionary tracks and related stellar parameters to estimate the rotational velocities of about 2000 B6- to F2-type normal stars. The velocity distribution of these normal stars is displayed in Figure 7 of \citet{2012A&A...537A.120Z} and the maximum peak value is at 150 $\text{km s}^{-1}$. Obviously, compared with normal A-type stars, our sample generally has lower rotational velocities.

\begin{figure}
	\centering
	\includegraphics[width=12cm]{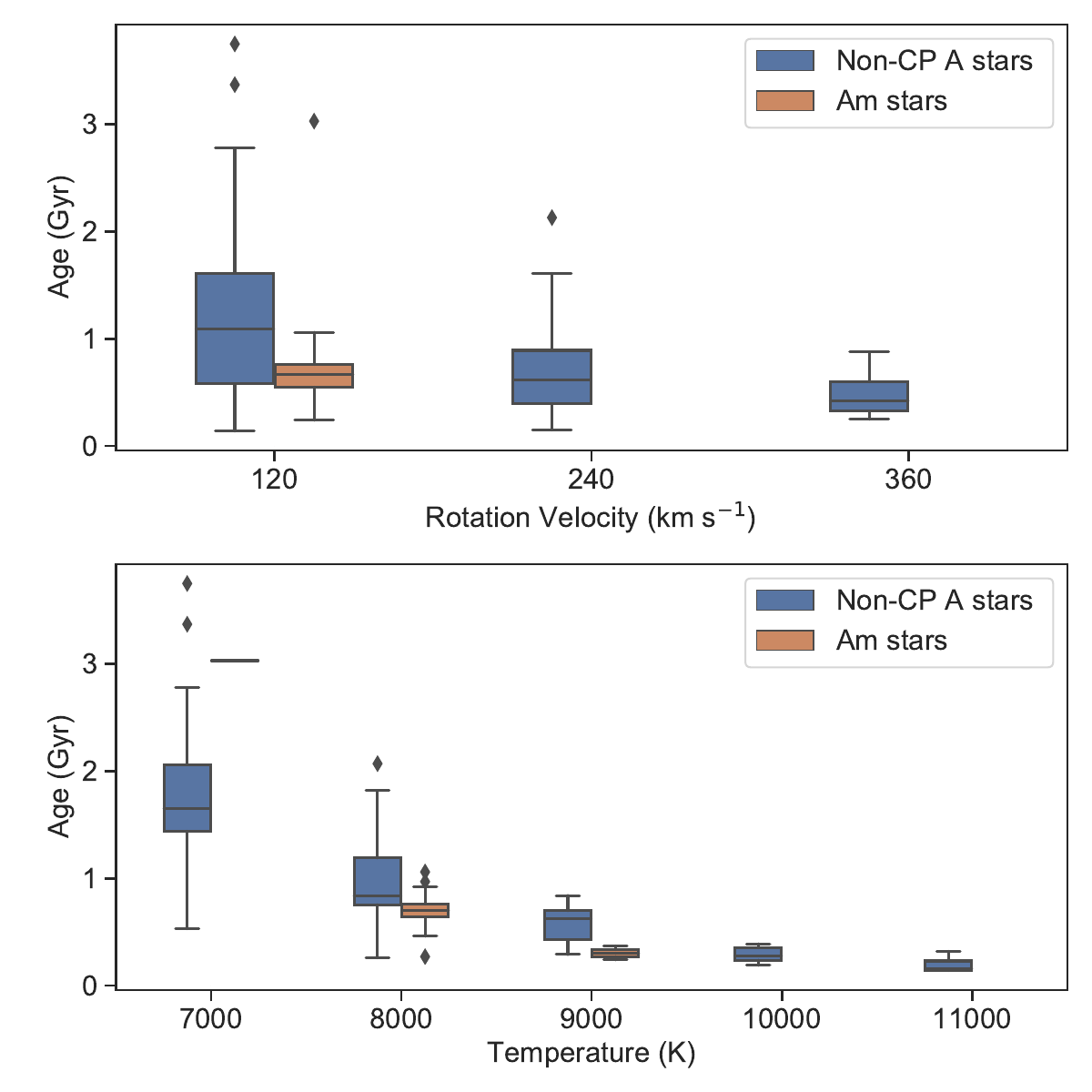}
	\caption{The top panel is rotation velocity vs. age. Stars were grouped according to rotation velocity, where  bin=$120\text{ km s}^{-1}$. The blue box represents the age distribution of each group of non-CP stars, and the red represents that of Am stars. Compared with the non-CP group with $V_{\rm rot} \leqslant 120\text{ km s}^{-1}$, the Am star group is younger. The bottom panel is the temperature vs. age for the stars with  $V_{\rm rot} \leqslant 120\text{ km s}^{-1}$. The color of the boxes is the same as the top panel. The average age of both Am and non-CP stars decreases with increasing temperature. Am stars are younger than non-CP stars in the same temperature group.}
\label{fig:age}
\end{figure}

We make use of the box plot to analyze the relationship among the rotation velocity, age, and temperature for non-CP and Am stars. In the top panel of Figure \ref{fig:age}, we divide the sample into three groups: $V_{\rm rot} \leqslant 120\text{ km s}^{-1}$, $120\text{ km s}^{-1} < V_{\rm rot} \leqslant 240\text{ km s}^{-1}$, and $ V_{\rm rot}  > 240\text{ km s}^{-1} $. Blue boxes represent non-CP stars and red boxes are Am stars. We find that the average age of the three groups of non-CP stars increases as rotation velocity decreases. Am stars are statistically younger than slowly rotational non-CP stars. We further analyzed the relationship of effective temperature and age of the stars with $V_{\rm rot} \leqslant 120\text{ km s}^{-1}$ in the bottom panel. Grouping according to temperature (bin = 1000 K), non-CP stars are divided into five groups, and Am stars are divided into two groups. The color of the boxes is the same as the top panel. From this subfigure, we find that as the temperature increases, the average ages of both non-CP and Am stars show a downward trend and the average age of non-CP stars is larger than that of Am stars in the same temperature group.

\subsection{Period and Amplitude}\label{sec:P_A}

\begin{figure}
	\centering
	\includegraphics[width=12cm]{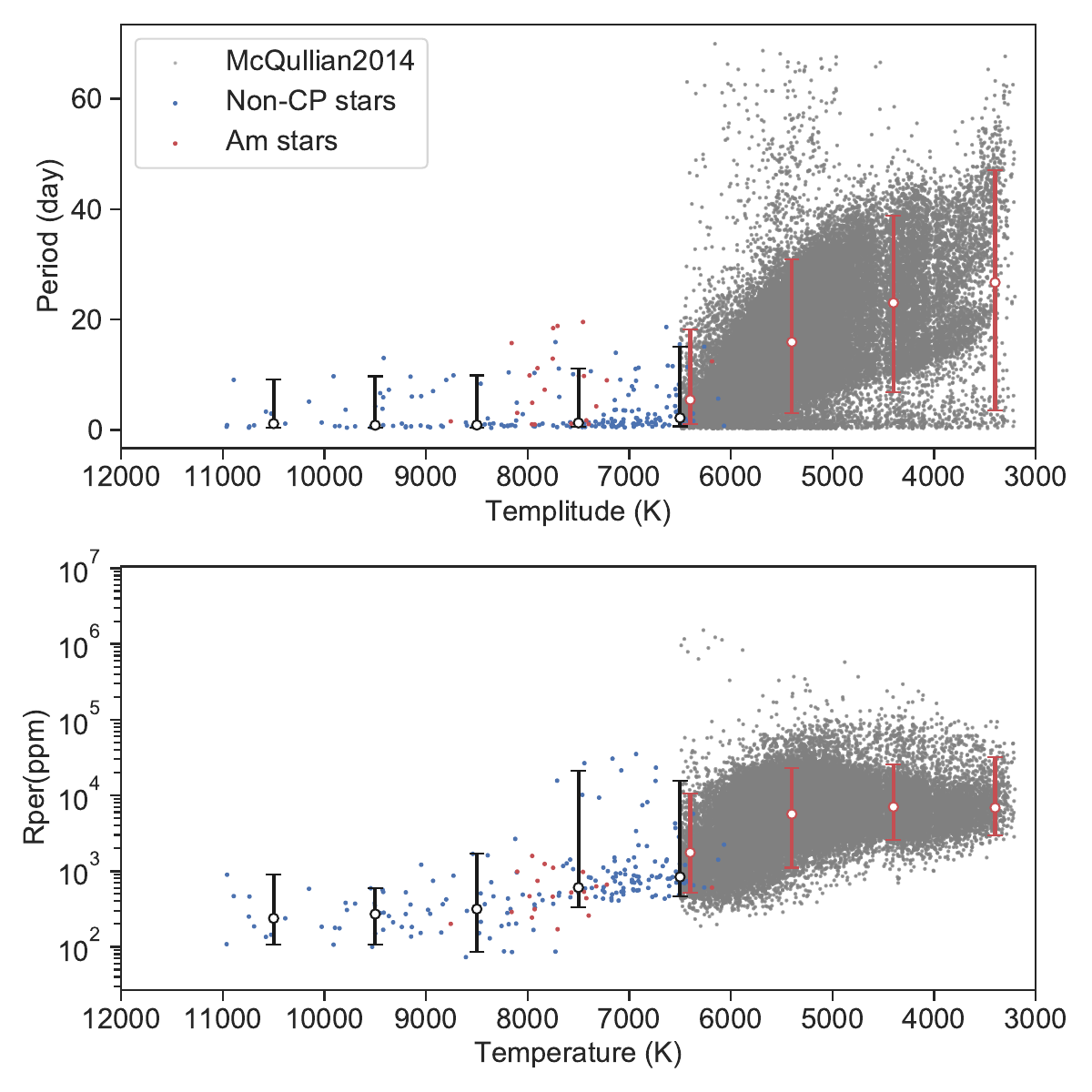}
	\caption{The distribution of Period and Rper in the temperature range of 3000K to 11,000K. Gray dots represent the data from \citet{2014ApJS..211...24M}. Blue and red dots is non-CP stars and Am stars in our sample. Rper is the average amplitude. Non-CP stars and cold stars are grouped by temperature (bin = 1000 K). The black (or red) hollow dots and error bars indicate the median value, 95\% value and 5\% value of non-CP stars (or cold stars) in each group. The median value of period and Rper increases as the temperature decreases in our sample. There are many non-CP stars for which the dispersion does not differ in either period or amplitude from that of Am stars between 7000 K to 8000 K.}
	\label{fig:teff_period_amplitude}
\end{figure}

\begin{figure}
	\centering
	\includegraphics[width=12cm]{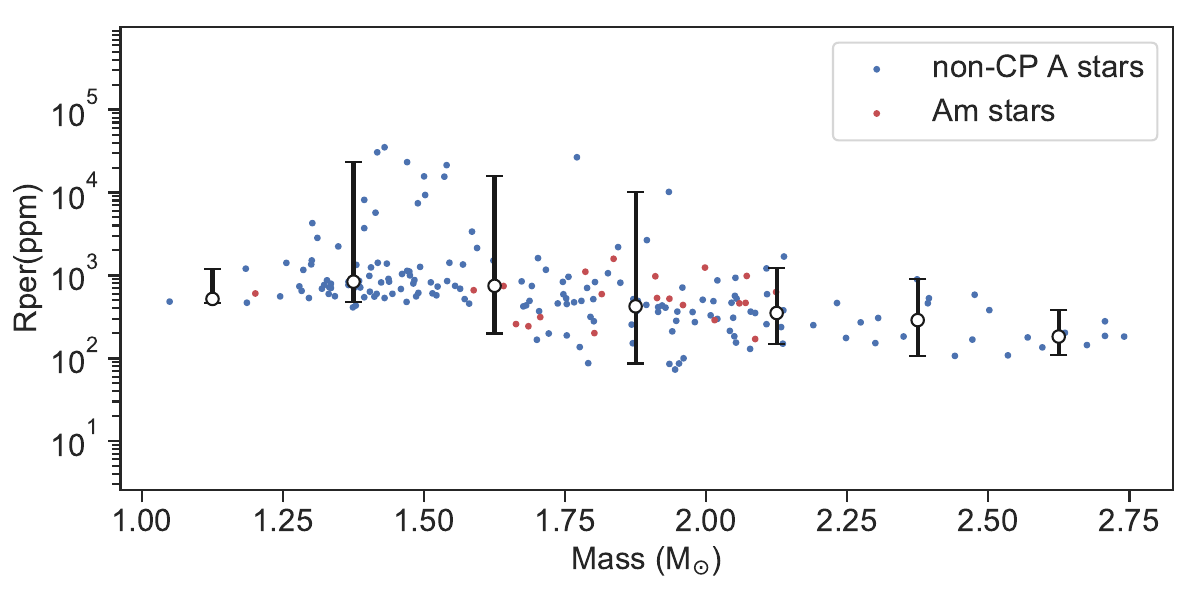}
	\caption{Amplitude vs. mass for the non-CP stars and the Am stars in our sample. Non-CP stars are grouped by stellar mass (bin = 0.25 M$_{\odot}$). The black hollow dots and error bars indicate the same of Figure \ref{fig:teff_period_amplitude}. The average amplitude decreases as the mass increases. }
	\label{fig:amplitude_mass}
\end{figure}

\citet{2014ApJS..211...24M} utilized the autoACF method to look for rotation periods of the MS stars with $T_{\rm eff} < $6500 K in Kepler data and analyzed the relationship among the period, the average amplitude, and the temperature. On this basis, we supplement the data of early A-type stars and provide a complete period and amplitude distribution over a larger temperature range, shown in Figure \ref{fig:teff_period_amplitude}. The data of \citet{2014ApJS..211...24M} are marked with gray dots, and the non-CP stars and Am stars in our sample are blue and red dots. The black or red hollow points with error bars indicate the median, 95\%, and 5\% values of non-CP stars or cold stars in each temperature group (bin = 1000 K).  

In the top panel of Figure \ref{fig:teff_period_amplitude}, the rotational period still obeys the law of cold stars that the rotation period becomes larger as the temperature decreases. It is just that the upward trend is slower. In the bottom panel, the vertical axis Rper represents the average photometric amplitude. We divide the light curve into many segments according to the rotation period. Rper is the average $R_{\text{var}}$ value of all segments (the interval between the 5th and 95th percentiles of the flux of each segment). The distribution of amplitude in non-CP stars is between 129 ppm ( the 5th percentile) and 7413 ppm ( the 95th percentile), with a median of 560 ppm. Compared with the results of \citet{2014ApJS..211...24M} (950, 22700, and 5600 ppm), they are all an order of magnitude lower. And the amplitude increases with decreasing temperature in our sample.

Am stars are mainly distributed in the temperature range of 7000K to 8000K. In this interval, the degree of dispersion of red dots in Figure \ref{fig:teff_period_amplitude} is equivalent to that of blue dots. When non-CP stars with a rotation velocity greater than 120 $\text{km s}^{-1}$ are removed from our sample, the dispersion of low-velocity rotational non-CP stars in Figure \ref{fig:teff_period_amplitude} remains basically unchanged. Therefore, we believe that there is no difference in either the rotational period or the average amplitude between Am and non-CP stars. This is consistent with the conclusion of \citet{2020MNRAS.492.3143T}, which suggests that Am and non-CP stars may have similar surface inhomogeneity.

Using stellar evolution calculations and an equipartition dynamo, \citet{2019ApJ...883..106C} theoretically prove there are dynamo-generated magnetic fields of order a few Gauss on the surface of most A-type stars and predicts that the related photometric variability will decrease as the stellar mass and effective temperature increase. Figure \ref{fig:amplitude_mass} is the distribution of Rper in the stellar mass of our sample. Combined with the analysis results of the bottom panel of Figure \ref{fig:teff_period_amplitude}, the change trends of Rper are consistent with the prediction of \citet{2019ApJ...883..106C}.

\subsection{Flare}\label{subsec:disc}
Through the long-term observation of the Sun, it is found that solar flares often occur above the sunspot groups. It is reasonable to speculate that similar conditions exist in solar-like stars. Whether there are flares in A-type stars is a topic of constant debate. \citet{2012MNRAS.423.3420B} and \citet{2013MNRAS.431.2240B} detected 33 A-type stars with flare events in the Kepler time-series photometry data. \citet{2017MNRAS.466.3060P} conducted in-depth investigations on 33 stars. They believed that these flares might not be inherent to the stars because most stars are contaminated by cold companion stars or cold background stars. \citet{2019ApJS..241...29} provided a flare catalog for the Kepler mission. This catalog contains A- to M-type flare stars. We cross match our sample with the catalog and find four common sources, all of which are non-CP stars.

\begin{figure}
	\centering
	\includegraphics[width=12cm]{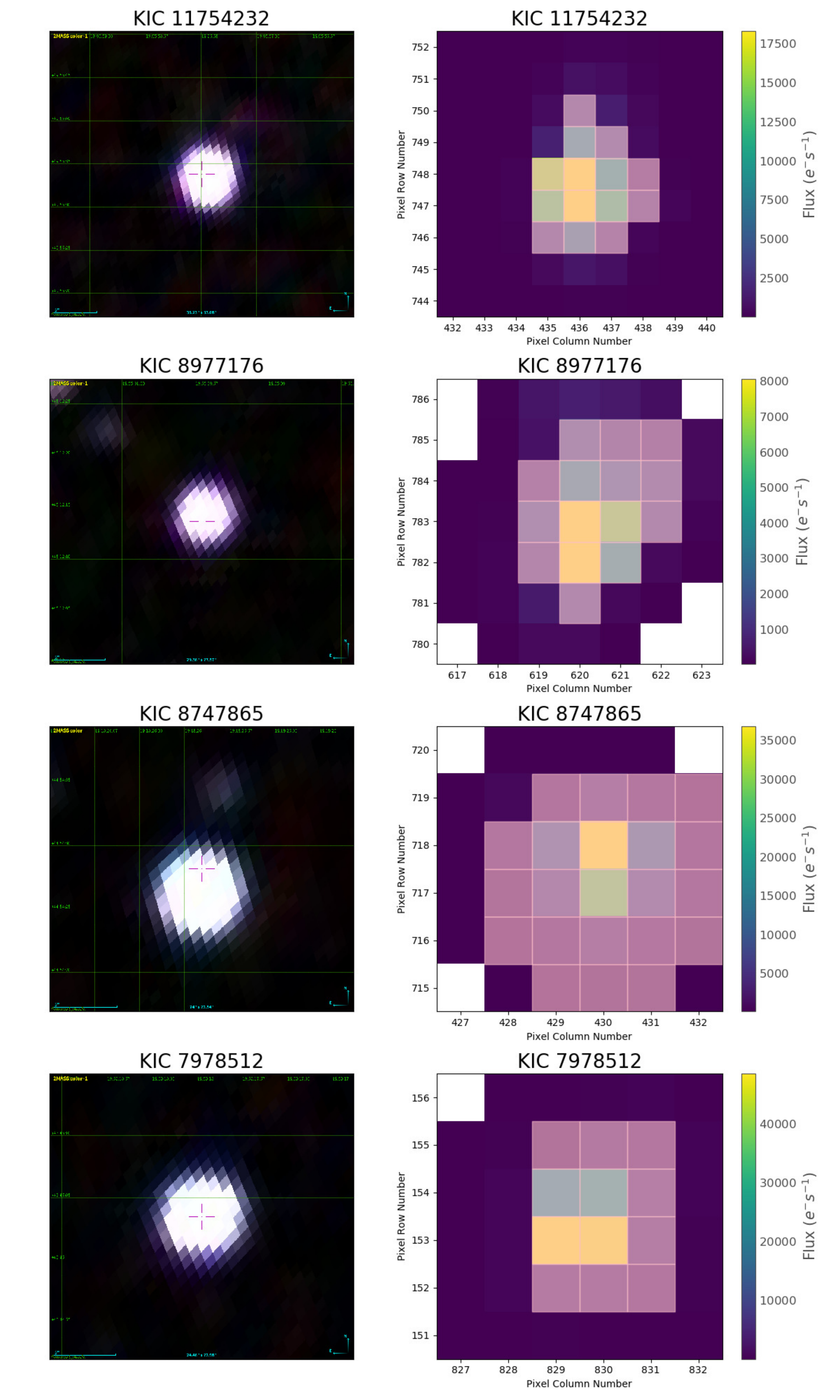}
	\caption{FOV images of 2MASS and pixel images of Kepler for KIC 11754232, KIC 8977176, KIC 8747865, and KIC 7978512. The light curve data are extracted from the pink mask. The color in the color bar represents the number of electrons received per second on the pixel.
	}
	\label{fig:view_field}
\end{figure}

We check the light curves of these four stars: KIC 11754232, KIC 8977176, KIC 8747865, and KIC 7978512, and detect that they all have multiple flare-like events.  We further inspect their surrounding field-of-view and pixel data during the flare period. We discover flare-like events in the pixel light curves of all four stars. Figure \ref{fig:view_field} is the FOV images from 2MASS and the pixel images of Kepler of these four stars. Because the Kepler telescope pixel resolution is 3.98$''$, we multiply the number of rows and columns of the pixel image by the pixel resolution to obtain the size of the FOV image. From the FOV images of 2MASS, we find other stars near the targets KIC 8977176 and KIC8747865. The pink area in the Kepler pixel image is the optimal mask, and the light curve is extracted from this area. Comparing the FOV image and the pixel image, stars near the target are not included in the optimal mask.

\begin{figure*}[ht]
	\centering
	\includegraphics[width=17cm]{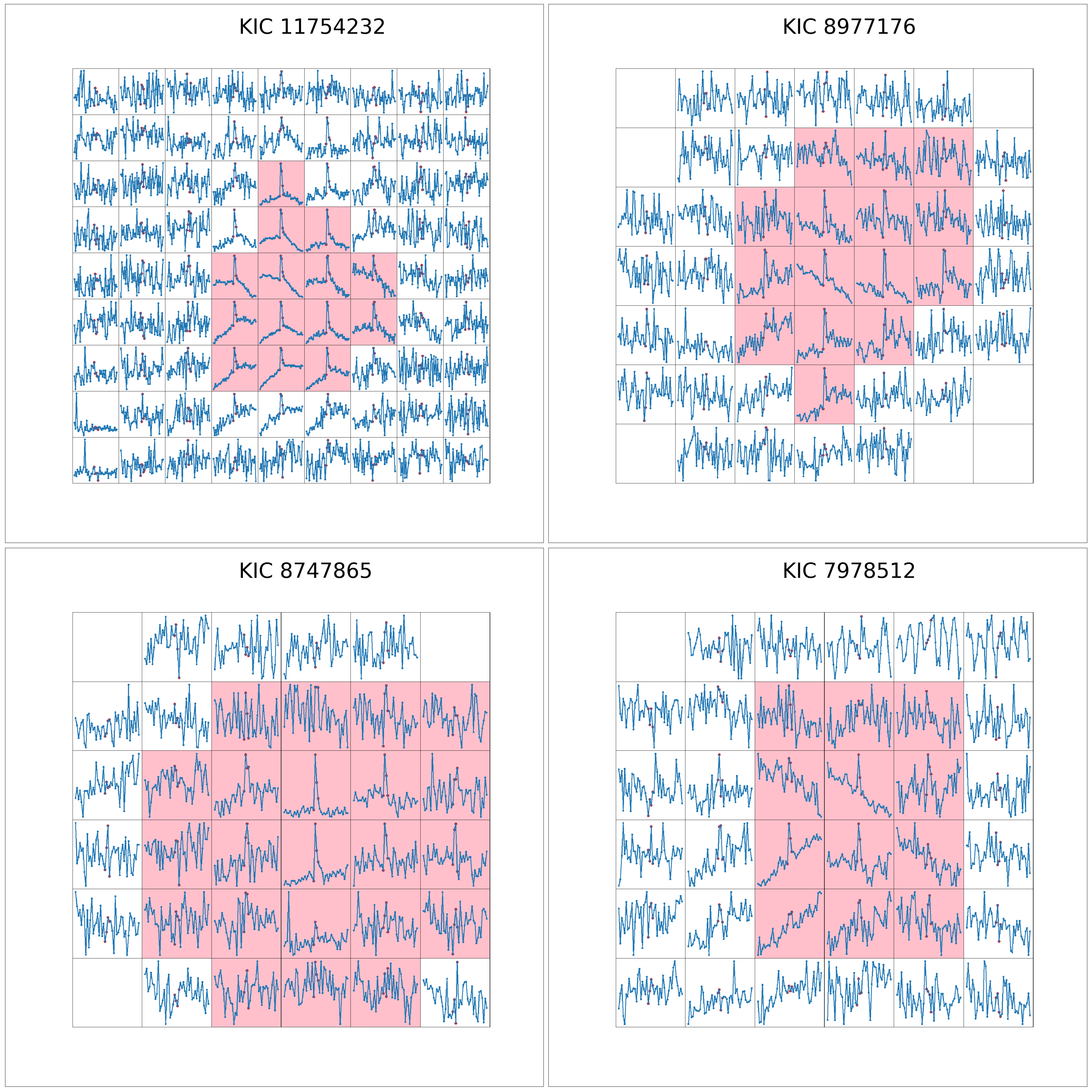}
	\caption{Pixel-level light curve with flare. Pink area are best aperture mask. Red dots represent the observation data during flare-events occurrence.}
	\label{fig:all_pixel}
\end{figure*}

Figure \ref{fig:all_pixel} shows the pixel light curves of the four targets during the flare period. Judging from pixel positions of the flares, we suspect that these flares come from the stars themselves. It is necessary to use a high-resolution spectrograph to make follow-up observations for these four stars to further check whether these flares are from companion stars.

\section{Conclusion}\label{sec:con}
In the LAMOST-Kepler project, we make use of LAMOST spectra to screen out 111 Am stars and 1154 non-CP stars and then utilize Kepler's light curve to calculate rotation velocity for 69 Am and 521 non-CP stars in the above sample. After removing the known binaries and pollution sources within four arcsec of the targets, we take advantage of the VC method to further eliminate the contamination of EBs. With the method of ACF, we check the rotation frequency manually. Using the projected rotational velocity, we further check the reliability of the rotational velocity. Finally, we obtain 199 stars with reliable rotation velocities, of which 21 are Am stars and 178 are non-CP stars.

The stars in our sample all have rotation modulation features. By analyzing the velocity distribution, we find that most of them are slower rotators. All Am stars have rotational velocity $ < $ 120 $\text{km s}^{-1} $, which is consistent with the results of previous studies. There are also a large number of slowly rotational non-CP stars in our sample, which bring light to a profound understanding of the CP stars, especially for Am stars. We divide non-CP stars into three groups: rotation velocity $\leqslant$ 120 $\text{km s}^{-1} $, 120 $\text{km s}^{-1} $ $<$ rotation velocity $\leqslant$ 240 $\text{km s}^{-1} $, and rotation velocity $>$ 240 $\text{km s}^{-1} $. And find that the average age of each group non-CP stars increases as the rotation velocity decreases. For the stars with rotation velocity $ < $ 120 $\text{km s}^{-1} $, the age of both non-CP stars and Am stars decreases with increasing temperature, and Am stars are statistically younger than non-CP stars in the same temperature bin.

We calculate the average photometric amplitude of Am stars and non-CP stars. Based on the results of \citet{2014ApJS..211...24M}, we provide a rotation period and amplitude distribution in a wider temperature range. We find that the rotation period and amplitude have an increasing trend as the temperature decreases for non-CP stars. There is no significant difference in either period or amplitude between non-CP stars and Am stars in the same temperature range, even after removing the high-velocity non-CP rotators, implying that non-CP stars and Am stars have similar surface inhomogeneities. We have also analyzed the relationship between amplitude and mass, and discover that the amplitude of non-CP stars basically decreases as the mass increases. Our analysis results are consistent with the theoretical predictions of \citet{2019ApJ...883..106C} about dynamo-generated magnetic fields in A-type stars, i.e. the related photometric variability will increase as the stellar mass and effective temperature decrease.

We find four stars with flare events in our sample, which are all non-CP stars. By examining FOV images, pixel images, and pixel-level light curves, the flare events in these four stars all tend to originate from the stars themselves, rather than from cold companion stars or cold background stars.

\section*{Acknowledgements}

We thank the anonymous referee for valuable comments to help us improve the paper. This work is supported by National Science Foundation of China (Nos U1931209, 12003050, 11973060, 11903008, U1631131), National Key R\&D Program of China(No. 2019YFA0405502), Key Research Program of Frontier Sciences, Chinese Academy of Sciences (No. QYZDY-SSW-SLH007), and Chinese Space Station Telescope (CSST) pre-research projects of Key Problems in Binaries and Chemical Evolution of the Milky Way and its Nearby Galaxies. Guoshoujing Telescope (the Large Sky Area Multi-Object Fiber Spectroscopic Telescope, LAMOST) is a National Major Scientific Project built by the Chinese Academy of Sciences. Funding for the project has been provided by the National Development and Reform Commission. LAMOST is operated and managed by the National Astronomical Observatories, Chinese Academy of Sciences. This paper includes data collected by the Kepler mission and obtained from the MAST data archive at the Space Telescope Science Institute (STScI). Funding for the Kepler mission is provided by the NASA Science Mission Directorate. STScI is operated by the Association of Universities for Research in Astronomy, Inc., under NASA contract NAS 5-26555. This research made use of Lightkurve, a Python package for Kepler and TESS data analysis \citep{2018ascl.soft12013L}.


\begin{thebibliography}{}
\expandafter\ifx\csname natexlab\endcsname\relax\def\natexlab#1{#1}\fi
\providecommand{\url}[1]{\href{#1}{#1}}
\providecommand{\dodoi}[1]{doi:~\href{http://doi.org/#1}{\nolinkurl{#1}}}
\providecommand{\doeprint}[1]{\href{http://ascl.net/#1}{\nolinkurl{http://ascl.net/#1}}}
\providecommand{\doarXiv}[1]{\href{https://arxiv.org/abs/#1}{\nolinkurl{https://arxiv.org/abs/#1}}}

\end{thebibliography}


\begin{thebibliography}{99}
\bibitem[Abt \& Levy(1985)]{1985ApJS...59..229A} Abt, H.~A. \& Levy, S.~G.\ 1985, \apjs, 59, 229. doi:10.1086/191070
\bibitem[Abt \& Morrell(1995)]{1995ApJS...99..135A} Abt, H.~A. \& Morrell, N.~I.\ 1995, \apjs, 99, 135. doi:10.1086/192182
\bibitem[Abt(2009)]{2009AJ....138...28A} Abt, H.~A.\ 2009, \aj, 138, 28. doi:10.1088/0004-6256/138/1/28
\bibitem[Balona(2011)]{2011MNRAS.415.1691B} Balona, L.~A.\ 2011, \mnras, 415, 1691. doi:10.1111/j.1365-2966.2011.18813.x
\bibitem[Balona(2012)]{2012MNRAS.423.3420B} Balona, L.~A.\ 2012, \mnras, 423, 3420. doi:10.1111/j.1365-2966.2012.21135.x
\bibitem[Balona et al.(2013)]{2013MNRAS.431.2240B} Balona, L.~A.\ 2013, \mnras, 431, 2240. doi:10.1093/mnras/stt322
\bibitem[Balona et al.(2015)]{2015MNRAS.448.1378B} Balona, L.~A., Catanzaro, G., Abedigamba, O.~P., et al.\ 2015, \mnras, 448, 1378. doi:10.1093/mnras/stv076
\bibitem[Balona et al.(2017)]{2017MNRAS.467.1830B} Balona, L.~A.\ 2017, \mnras, 467, 1830. doi:10.1093/mnras/stx265
\bibitem[Berger et al.(2020)]{2020AJ....159..280B} Berger, T.~A., Huber, D., van Saders, J.~L., et al.\ 2020, \aj, 159, 280. doi:10.3847/1538-3881/159/6/280
\bibitem[Blaz{\`e}re et al.(2016a)]{2016MNRAS.459L..81B} Blaz{\`e}re, A., Neiner, C., \& Petit, P.\ 2016, \mnras, 459, L81. doi:10.1093/mnrasl/slw050
\bibitem[Blaz{\`e}re et al.(2016b)]{2016A&A...586A..97B} Blaz{\`e}re, A., Petit, P., Ligni{\`e}res, F., et al.\ 2016, \aap, 586, A97. doi:10.1051/0004-6361/201527556
\bibitem[B{\"o}hm et al.(2015)]{2015A&A...577A..64B} B{\"o}hm, T., Holschneider, M., Ligni{\`e}res, F., et al.\ 2015, \aap, 577, A64. doi:10.1051/0004-6361/201425425
\bibitem[Cantiello \& Braithwaite(2019)]{2019ApJ...883..106C} Cantiello, M. \& Braithwaite, J.\ 2019, \apj, 883, 106. doi:10.3847/1538-4357/ab3924
\bibitem[Carquillat \& Prieur(2007)]{2007MNRAS.380.1064C} Carquillat, J.-M. \& Prieur, J.-L.\ 2007, \mnras, 380, 1064. doi:10.1111/j.1365-2966.2007.12143.x
\bibitem[Catanzaro \& Ripepi(2014)]{2014MNRAS.441.1669C} Catanzaro, G. \& Ripepi, V.\ 2014, \mnras, 441, 1669. doi:10.1093/mnras/stu674
\bibitem[Catanzaro et al.(2015)]{2015MNRAS.451..184C} Catanzaro, G., Ripepi, V., Biazzo, K., et al.\ 2015, \mnras, 451, 184. doi:10.1093/mnras/stv952
\bibitem[Conti(1965)]{1965ApJ...142.1594C} Conti, P.~S.\ 1965, \apj, 142, 1594. doi:10.1086/148439
\bibitem[Cui et al.(2012)]{2012RAA....12.1197C} Cui, X.-Q., Zhao, Y.-H., Chu, Y.-Q., et al.\ 2012, Research in Astronomy and Astrophysics, 12, 1197. doi:10.1088/1674-4527/12/9/003
\bibitem[De Cat et al.(2015)]{2015ApJS..220...19D} De Cat, P., Fu, J.~N., Ren, A.~B., et al.\ 2015, \apjs, 220, 19. doi:10.1088/0067-0049/220/1/19
\bibitem[Gray(1999)]{1999ascl.soft10002G} Gray, R.~O.\ 1999, Astrophysics Source Code Library. ascl:9910.002
\bibitem[Gray \& Corbally(2009)]{2009ssc..book.....G} Gray, R.~O. \& Corbally, C.\ 2009, Stellar Spectral Classification by Richard O. Gray and Christopher J. Corbally. Princeton University Press, 2009. ISBN: {\^A} 978-0-691-12511-4
\bibitem[Gray et al.(2016)]{2016AJ....151...13G} Gray, R.~O., Corbally, C.~J., De Cat, P., et al.\ 2016, \aj, 151, 13. doi:10.3847/0004-6256/151/1/13
\bibitem[Gilliland et al.(2015)]{2015AJ....150..133G} Gilliland, R.~L., Chaplin, W.~J., Jenkins, J.~M., et al.\ 2015, \aj, 150, 133. doi:10.1088/0004-6256/150/4/133
\bibitem[Yang et al.(2019)]{2019ApJS..241...29} Yang, H., Liu, J., \ 2019, \apjs, 241, 29. doi:10.3847/1538-4365/ab0d28
\bibitem[H{\"u}mmerich et al.(2020)]{2020A&A...640A..40H} H{\"u}mmerich, S., Paunzen, E., \& Bernhard, K.\ 2020, \aap, 640, A40. doi:10.1051/0004-6361/202037750
\bibitem[Kirk et al.(2016)]{2016AJ....151...68K} Kirk, B., Conroy, K., Pr{\v{s}}a, A., et al.\ 2016, \aj, 151, 68. doi:10.3847/0004-6256/151/3/68
\bibitem[Lane \& Lester(1987)]{1987ApJS...65..137L} Lane, M.~C. \& Lester, J.~B.\ 1987, \apjs, 65, 137. doi:10.1086/191220
\bibitem[Lightkurve Collaboration et al.(2018)]{2018ascl.soft12013L} Lightkurve Collaboration, Cardoso, J.~V. de M., Hedges, C., et al.\ 2018, Astrophysics Source Code Library. ascl:1812.013
\bibitem[Ligni{\`e}res et al.(2009)]{2009A&A...500L..41L} Ligni{\`e}res, F., Petit, P., B{\"o}hm, T., et al.\ 2009, \aap, 500, L41. doi:10.1051/0004-6361/200911996
\bibitem[McQuillan et al.(2013)]{2013MNRAS.432.1203M} McQuillan, A., Aigrain, S., \& Mazeh, T.\ 2013, \mnras, 432, 1203. doi:10.1093/mnras/stt536
\bibitem[McQuillan et al.(2014)]{2014ApJS..211...24M} McQuillan, A., Mazeh, T., \& Aigrain, S.\ 2014, \apjs, 211, 24. doi:10.1088/0067-0049/211/2/24
\bibitem[Pedersen et al.(2017)]{2017MNRAS.466.3060P} Pedersen, M.~G., Antoci, V., Korhonen, H., et al.\ 2017, \mnras, 466, 3060. doi:10.1093/mnras/stw3226
\bibitem[Petit et al.(2011)]{2011A&A...532L..13P} Petit, P., Ligni{\`e}res, F., Auri{\`e}re, M., et al.\ 2011, \aap, 532, L13. doi:10.1051/0004-6361/201117573
\bibitem[Petit et al.(2014)]{2014A&A...568C...2P} Petit, P., Ligni{\`e}res, F., Wade, G.~A., et al.\ 2014, \aap, 568, C2. doi:10.1051/0004-6361/201015307e
\bibitem[Petit et al.(2017)]{2017MNRAS.472L..30P} Petit, P., H{\'e}brard, E.~M., B{\"o}hm, T., et al.\ 2017, \mnras, 472, L30. doi:10.1093/mnrasl/slx132
\bibitem[Qin et al.(2019)]{2019ApJS..242...13Q} Qin, L., Luo, A.-L., Hou, W., et al.\ 2019, \apjs, 242, 13. doi:10.3847/1538-4365/ab17d8
\bibitem[Saio et al.(2018)]{2018MNRAS.474.2774S} Saio, H., Kurtz, D.~W., Murphy, S.~J., et al.\ 2018, \mnras, 474, 2774. doi:10.1093/mnras/stx2962
\bibitem[Savanov(2018)]{2018ARep...62..814S} Savanov, I.~S.\ 2018, Astronomy Reports, 62, 814. doi:10.1134/S1063772918110057
\bibitem[Sikora et al.(2019a)]{2019MNRAS.483.2300S} Sikora, J., Wade, G.~A., Power, J., et al.\ 2019a, \mnras, 483, 2300. doi:10.1093/mnras/sty3105
\bibitem[Sikora et al.(2019b)]{2019MNRAS.487.4695S} Sikora, J., David-Uraz, A., Chowdhury, S., et al.\ 2019b, \mnras, 487, 4695. doi:10.1093/mnras/stz1581
\bibitem[Sikora et al.(2020)]{2020MNRAS.498.2456S} Sikora, J., Wade, G.~A., \& Rowe, J.\ 2020, \mnras, 498, 2456. doi:10.1093/mnras/staa2444
\bibitem[Smalley et al.(2014)]{2014A&A...564A..69S} Smalley, B., Southworth, J., Pintado, O.~I., et al.\ 2014, \aap, 564, A69. doi:10.1051/0004-6361/201323158
\bibitem[Smith(1996)]{1996Ap&SS.237...77S} Smith, K.~C.\ 1996, \apss, 237, 77. doi:10.1007/BF02424427
\bibitem[Tian et al.(2020)]{2020ApJS..249...22T} Tian, Z., Liu, X., Yuan, H., et al.\ 2020, \apjs, 249, 22. doi:10.3847/1538-4365/ab9904
\bibitem[Trust et al.(2020)]{2020MNRAS.492.3143T} Trust, O., Jurua, E., De Cat, P., et al.\ 2020, \mnras, 492, 3143. doi:10.1093/mnras/stz3623
\bibitem[VanderPlas(2018)]{2018ApJS..236...16V} VanderPlas, J.~T.\ 2018, \apjs, 236, 16. doi:10.3847/1538-4365/aab766
\bibitem[Zhao et al.(2012)]{2012RAA....12..723Z} Zhao, G., Zhao, Y.-H., Chu, Y.-Q., et al.\ 2012, Research in Astronomy and Astrophysics, 12, 723. doi:10.1088/1674-4527/12/7/002
\bibitem[Zorec \& Royer(2012)]{2012A&A...537A.120Z} Zorec, J. \& Royer, F.\ 2012, \aap, 537, A120. doi:10.1051/0004-6361/201117691
\end{thebibliography}
\end{document}